\documentstyle[12pt]{article}
\begin{document}
\begin{center}
{\Large \bf GLUON CONTRIBUTION TO THE TRANSVERSE
SPIN STRUCTURE FUNCTION $g_2$}\\
\large
{A.V.Belitsky, A.V.Efremov, O.V.Teryaev}\\[0.3cm]
{\it Bogolubov Laboratory of Theoretical Physics\\
Joint Institute for Nuclear Research\\
141980, Dubna, Russia}
\end{center}
\normalsize
\begin{center}
\large Abstract
\end{center}
\large
We consider the twist-3 gluon contribution to the transverse spin structure
function $g_2$.We find that the Burkhardt-Cottingham sum rule is satisfied.
The additional {\it nonlocal} operators appear
in the moments of $g_2$ which were absent in previous analysis of
Bukhvostov, Kuraev and Lipatov.

\newpage
\large
The only way we can get the information about the nucleon spin content
in the polarized deep inelastic experiment is to measure the
structure functions $g_1$ and $g_2$ which are the coefficients in the
decomposition of imaginary part of the forward Compton
($T_{\alpha \beta}$) amplitude over
appropriate tensor structures:
\begin{eqnarray}
&&\!\!\!\!\! W_{[\alpha \beta]}={1\over {\pi }}ImT_{[\alpha \beta]}
\nonumber\\
&&\!\!\!\!\! ={i\over {(pq)}}{\epsilon}_{\alpha \beta \rho \sigma}
q_{\rho}\Bigl[ g_1(x_B,Q^2)s_{\sigma } +
g_2(x_B,Q^2)\biggl(s_{\sigma }-p_{\sigma }{(sq)\over (pq)}\biggr)\Bigr]
\end{eqnarray}
and
\begin{equation}
T_{\alpha \beta}=i\int d^4xe^{iqx}\langle p,s|T(j_{\beta}(x)
j_{\alpha}(0))|p,s\rangle ,
\end{equation}
where $s_{\alpha}$ is a nucleon covariant polarization ( $s^2=-M^2$ for pure
state, $M$ is a nucleon mass ), $x_B={Q^2\over{ 2(pq)}}$ is a Bjorken
variable and $q_{\rho}$ is four-momentum transfer.

There exist a set of sum rules for both $g_1$ and $g_2$. We remind
the reader the Bjorken \cite{1}, Ellis-Jaffe \cite{2} sum rules
for the first one. The recent EMC result has put a great attention
to the first moment of $g_1$ due to the large discrepancy with
the na\"\i ve parton picture \cite{3}. As for the second polarized
structure function $g_2$ that measures the difference between
the property of transversely and longitudinally polarized nucleon
there exists the superconverging Burkhardt-Cottingham sum rule \cite{4}:
\begin{equation}
\int \limits_{0}^{1}dx_Bg_2(x_B)=0 .
\end{equation}
There is some discussion in the literature concerning the
validity of the last one \cite{5}. This issue is made particularly
important by near future measurement of $g_2$. Since the check of the
Burkhardt-Cottingham sum rule is one of the planned experimental task
it is essential to know whether the RHS of eq.(3) is expected
to be nonzero.

The papers on higher twist effects in polarized
deep inelastic scattering can be devided into three parts: the
first one involves the works where the authors consider only
flavour nonsinglet operators \cite{6}, in the second one the
singlet channel is included but the triple gluon component is
disregarded \cite{7} and in the last one but not the least the evolution
equations for partonic correlators is written where the twist-3 gluon
operators
appear in the problem of mixing with the other operators under
the renormalization group evolution \cite{8}. So, the anomalous
dimentions of gluon operators have already been found, however,
their coefficient functions have not, though they look more
important in practice. The gluons play the crucial role in
the quark confinement and therefore we expect the 3-gluon effect to
be large in the realistic nucleon. In this paper we investigate
this question and present some physical implication of our
results.

In the following discussion we will work in the light-cone gauge,
$B_+\sim (Bn)=0$, where the vector $n_{\mu}$ can be constructed from
the vectors
of the problem: $n_{\mu}=\frac{1}{(pq)}[q_{\mu}+x_Bp_{\mu}]$. One of the
advantages of it is that one can easily recover
gauge invariant expression after all required calculations have done.

The starting point of the analysis is the factorization procedure
developed in the Ref. \cite{9} which results in the separation of the
hard coefficient function $(E)$ from soft correlation function $(N)$.
On the first stage of factorization we obtain gauge variant formula
for correlator in terms of gluon fields
$\sim \langle B_{\mu}B_{\nu}B_{\sigma} \rangle$.
Making use of the equation which directly connects gluon strength
tensor and field itself ( we drop the possible constant term due to
the arbitrariness in the boundary condition ):
\begin{equation}
B_{\mu}^a(\lambda n)=\frac{1}{2}\int_{-\infty}^{\infty}dz \epsilon
(\lambda n^{-}-z)G_{+\mu}^a(z),
\end{equation}
we acquire additional factor $[x_1x_2(x_2-x_1)]^{-1}$ in the factorized
expression for the structure function.

The three-gluon correlation function\footnote{\large Because of C-even
photon state in the t-channel there appears only $f$-convolution
of gluon fields. } can be decomposed into the two
independent scalar functions time appropriate tensor structures \cite{10}:
\begin{eqnarray}
&\!\!\!\! N_{\mu \nu \sigma}(x_1,x_2) \nonumber\\
&\!\!\! =\!-g(in^{-})^3\!\int\!\!\frac{d\lambda }{2\pi }\frac{d\mu }{2\pi }
e^{-i\lambda x_2}e^{-i\mu (x_1-x_2)}\langle p,s|if^{abc}G_{+\mu}^a(0)
G_{+\nu}^b(\mu n)G_{+\sigma}^c(\lambda n)|p,s\rangle  \nonumber\\
&\!\!\!\! =2i[N(x_1,x_2)g_{\sigma \mu}{\epsilon}_{\nu pns}
-N(x_1,x_1-x_2)g_{\mu \nu}{\epsilon}_{\sigma pns}-N(x_2,x_2-x_1)
g_{\nu \sigma}{\epsilon}_{\mu pns} \nonumber\\
&\!\!\!\! +\widetilde N(x_1,x_2){\epsilon}_{\sigma \mu pn}s_{\nu}
+\widetilde N(x_1,x_1-x_2){\epsilon }_{\mu \nu pn}s_{\sigma}
-\widetilde N(x_2,x_2-x_1){\epsilon }_{\nu \sigma pn}s_{\mu}] .
\end{eqnarray}
The hermiticity leads to the following symmetry properties of $N$
and $\widetilde N$:
\begin{eqnarray}
N(x_1,x_2)=N(x_2,x_1)&,&N(x_1,x_2)=-N(-x_1,-x_2) ,\nonumber\\
\widetilde N(x_1,x_2)=-\widetilde N(x_2,x_1)&,&\widetilde N(x_1,x_2)
=-\widetilde N(-x_1,-x_2) .
\end{eqnarray}
An explicit form of $N$ and $\widetilde N$ functions in terms of
three-gluon correlators
can be found by projecting the LHS of eq.(4) onto the tensor structures
of the RHS. Solving the obtained six-by-six matrix equation with respect
to $N(x_1,x_2)$ and $\widetilde N(x_1,x_2)$ we come to the expressions:
\begin{eqnarray}
&\!\!\!\! 2s_{\perp}^2N(x_1,x_2)\nonumber\\
&\!\!\!\! =\!-g (in^{-})^3is_{\perp}^{\nu}\!\int\!\!\frac{d\lambda }{2\pi }
\frac{d\mu }{2\pi }e^{-i{\lambda}x_2}e^{-i{\mu}(x_1-x_2)}if^{abc}
\langle p,s| \bigl(\frac{1}{2}G_{+\mu }^a(0)\widetilde G_{+\nu }^b(\mu n)
G_{+\mu }^c(\lambda n)\nonumber\\
&\!\!\!\! -\frac{1}{6}G_{+\mu }^a(0)G_{+\mu }^b(\mu n)\widetilde
G_{+\nu }^c(\lambda n)-\frac{1}{6}\widetilde G_{+\nu }^a(0)G_{+\mu }^b
(\mu n)G_{+\mu }^c(\lambda n)\nonumber\\
&\!\!\!\! -\frac{1}{3}G_{+\mu }^a(0)\widetilde G_{+\mu }^b(\mu n)
G_{+\nu }^c(\lambda n)+\frac{1}{3}G_{+\nu }^a(0)G_{+\mu }^b(\mu n)
\widetilde G_{+\mu }^c(\lambda n)\bigr)|p,s\rangle
\end{eqnarray}
and
\begin{eqnarray}
&\!\!\!\! 2s_{\perp}^2\widetilde N(x_1,x_2)\nonumber\\
&\!\!\!\! =\!-g (in^{-})^3is_{\perp}^{\nu}\!\int\!\!\frac{d\lambda }
{2\pi }\frac{d\mu }{2\pi }e^{-i{\lambda}x_2}e^{-i{\mu}(x_1-x_2)}if^{abc}
\langle p,s| \bigl(\frac{1}{3}\widetilde G_{+\nu }^a(0)G_{+\mu }^b(\mu n)
G_{+\mu }^c(\lambda n)\nonumber\\
&\!\!\!\! -\frac{1}{3}G_{+\mu }^a(0)G_{+\mu }^b(\mu n)\widetilde
G_{+\nu }^c(\lambda n)+\frac{1}{3}G_{+\mu }^a(0)G_{+\nu }^b(\mu n)
\widetilde G_{+\mu }^c(\lambda n) \nonumber\\
&\!\!\!\! +\frac{2}{3}G_{+\mu }^a(0)\widetilde G_{+\mu }^b(\mu n)
G_{+\nu }^c(\lambda n)+\frac{2}{3}G_{+\nu }^a(0)G_{+\mu }^b(\mu n)
\widetilde G_{+\mu }^c(\lambda n)\bigr)|p,s\rangle .
\end{eqnarray}
As can be easily understood the first line of eq.(5) contributes to
the spin structure function $g_2$ while  the second one gives rise
to $g_2$ as well as to $g_1$. The last statement can be checked by
explicit convolution of eq.(4) and the general tensor decomposition
of the coefficient function:
\begin{eqnarray}
&t^{[\alpha \beta] \mu \nu \sigma }
=t^{(1)}_1{\epsilon}^{\alpha \beta q \nu}{\epsilon }^{\sigma \mu pq}
+t^{(1)}_2{\epsilon}^{\alpha \beta q}_{\hspace{6mm} \rho}g^{\sigma \mu}
{\epsilon }^{\nu pq \rho }
+t^{(1)}_3{\epsilon}^{\alpha \beta q}_{\hspace{6mm} \rho}p^{\nu}
{\epsilon }^{\sigma \mu q\rho }\nonumber\\
&+\frac{1}{(pq)}t^{(1)}_4{\epsilon}^{\alpha \beta qp}q^{\nu}
{\epsilon }^{\sigma \mu pq}
+\frac{(pq)}{q^2}t^{(1)}_5{\epsilon}^{\alpha \beta q}_{\hspace{6mm} \rho}
q^{\nu}{\epsilon }^{\sigma \mu q\rho }
+(\mu \rightarrow \nu \rightarrow \sigma \rightarrow \mu) .
\end{eqnarray}

In order to find the coefficient function we have to evaluate the
five-point Green function:
\begin{eqnarray}
&&\!\!\!\!\!\!\!\!t^{[\alpha \beta] \mu \nu \sigma } \nonumber\\
&&\!\!\!\!\!\!\!\!=\!\!\int\!\!d^4xe^{iqx}\langle 0|T(j_{[ \beta}(x)
j_{\alpha]}(0)if^{abc}B_{\mu}^a(x_1 p)B_{\nu}^b((x_2-x_1) p)
B_{\sigma}^c(-x_2 p))|0\rangle_{amp} ,\nonumber\\
\end{eqnarray}
whose imaginary part gives us the coefficient functions of
interest up to normalization factor that can be found by applying Wick
theorem to the forward Compton amplitude.
The relevant loop diagrams are shown in fig.1 (a,b).

The calculations are lengthy but straightforward and the obtained
results are:
\begin{eqnarray}
&&\!\!\!\!\!g_2^{3gluon}(x_B)=3\langle {Q_q}^2\rangle
N_f\frac{{\alpha}_s}{4\pi } \nonumber\\
&&\!\!\!\!\!\times \int \frac{dx_1dx_2}{x_1x_2(x_2-x_1)}
\left(E_2(x_1,x_2,x_B)N(x_1,x_2)+\widetilde E_2(x_1,x_2,x_B)
\widetilde N(x_1,x_2)\right) \nonumber\\
\end{eqnarray}
and
\begin{equation}
g_1^{3gluon}(x_B)=3\langle {Q_q}^2\rangle N_f\frac{{\alpha}_s}
{4\pi }\int \frac{dx_1dx_2}{x_1x_2(x_2-x_1)}\widetilde E_1(x_1,x_2,x_B)
\widetilde N(x_1,x_2) ,
\end{equation}
where $\langle {Q_q}^2\rangle=\sum_{}^{} \frac{Q_q^2}{N_f}$ is an
average quark charge squared in the elementary units. Performing
the calculation
%we take into account $SU(3)_{flavor}$ symmetry
%for light quarks: $m_u=m_d=m_s \equiv m_q$.
we have adopted the
quark mass $m_q$ as a IR-cutoff parameter:
\begin{eqnarray}
&&E_2(x_1,x_2,x_B)\nonumber\\
&&=\theta (x_B)\biggl[\frac{3(x_2+x_1)(x_B-x_2+x_1)}{x_1x_2(x_2-x_1)}
\theta (x_2-x_1-x_B)\nonumber\\
&&-\frac{(x_B-x_2)(3x_2-2x_1)}{x_1x_2^2}\theta (x_2-x_B)\nonumber\\
&&-\frac{(x_B-x_1)(2x_2-3x_1)}{x_1^2x_2}\theta (x_1-x_B)\nonumber\\
&&-\frac{(x_2+x_1)(x_B-x_2+x_1)}{x_1x_2(x_2-x_1)}ln\bigl(
\frac{x_2-x_1-x_B}{x_B}\bigr)\theta (x_2-x_1-x_B) \nonumber\\
&&+\frac{(x_2-x_1)(x_B-x_2)-x_Bx_1}{x_1x_2^2}ln\bigl(\frac{x_2-x_B}
{x_B}\bigr)\theta (x_2-x_B) \nonumber\\
&&+\frac{(x_2-x_1)(x_B-x_2)+x_Bx_2}{x_1^2x_2}ln\bigl(\frac{x_1-x_B}
{x_B}\bigr)\theta (x_1-x_B) \nonumber\\
&&+ln\left(\frac{m_q^2}{Q^2}\right)\biggl(\frac{(x_2+x_1)(x_B-x_2+x_1)}
{x_1x_2(x_2-x_1)}\theta (x_2-x_1-x_B) \nonumber\\
&&-\frac{(x_2-x_1)(x_B-x_2)-x_Bx_1}{x_1x_2^2}\theta (x_2-x_B) \nonumber\\
&&-\frac{(x_2-x_1)(x_B-x_2)+x_Bx_2}{x_1^2x_2}\theta (x_1-x_B)\biggr)
+(x_i\rightarrow -x_i)\biggr]\nonumber\\
&&+(x_B\rightarrow -x_B)
\end{eqnarray}
and
\begin{eqnarray}
&&\widetilde E_2(x_1,x_2,x_B)\nonumber\\
&&=\theta (x_B)\biggl[-\frac{(x_B-x_2+x_1)(3x_1^2
+3x_2^2-2x_1x_2)}{x_1x_2(x_2-x_1)^2}\theta (x_2-x_1-x_B)\nonumber\\
&&-\frac{(x_B-x_2)[18x_1x_2-(x_2-x_1)(3x_2+2x_1)]+4x_Bx_1x_2}
{x_1x_2^2(x_2-x_1)}\theta (x_2-x_B)\nonumber\\
&&+\frac{(x_B-x_1)[18x_1x_2+(x_2-x_1)(3x_1+2x_2)]+4x_Bx_1x_2}
{x_2x_1^2(x_2-x_1)}\theta (x_1-x_B)\nonumber\\
&&+\frac{(x_B-x_2+x_1)(x_2^2+x_1^2)+2x_1x_2}{x_1x_2(x_2-x_1)}
ln\bigl(\frac{x_2-x_1-x_B}{x_B}\bigr)\theta (x_2-x_1-x_B) \nonumber\\
&&+\frac{(x_B-x_2)[6x_1x_2-(x_2-x_1)(x_2+x_1)]+x_Bx_1(3x_2+x_1)}
{x_1x_2^2(x_2-x_1)}\nonumber\\
&&\hspace{8cm}\times ln\bigl(\frac{x_2-x_B}{x_B}\bigr)\theta (x_2-x_B)
\nonumber\\
&&-\frac{(x_B-x_1)[6x_1x_2+(x_2-x_1)(x_2+x_1)]+x_Bx_2(3x_1+x_2)}
{x_2x_1^2(x_2-x_1)}\nonumber\\
&&\hspace{8cm}\times ln\bigl(\frac{x_1-x_B}{x_B}\bigr)\theta (x_1-x_B)
\nonumber\\
&&+ln\left(\frac{m_q^2}{Q^2}\right)\biggl(-\frac{(x_B-x_2+x_1)
(x_2^2+x_1^2)+2x_1x_2}{x_1x_2(x_2-x_1)}\theta (x_2-x_1-x_B)
\nonumber\\
&&-\frac{(x_B-x_2)[6x_1x_2-(x_2-x_1)(x_2+x_1)]+x_Bx_1(3x_2+x_1)}
{x_1x_2^2(x_2-x_1)}\theta (x_2-x_B) \nonumber\\
&&+\frac{(x_B-x_1)[6x_1x_2+(x_2-x_1)(x_2+x_1)]+x_Bx_2(3x_1+x_2)}
{x_2x_1^2(x_2-x_1)}\theta (x_1-x_B) \biggr) \nonumber\\
&&+(x_i\rightarrow -x_i)\biggr]+(x_B\rightarrow -x_B) .
\end{eqnarray}
The coefficient function which enter the polarized structure function
$g_1$ is the following:
\begin{eqnarray}
&&\widetilde E_1(x_1,x_2,x_B)\nonumber\\
&&=\theta (x_B)\biggl[-4\frac{4x_B-3x_2}{x_2(x_2-x_1)}\theta (x_2-x_B)
+4\frac{4x_B-3x_1}{x_1(x_2-x_1)}\theta (x_1-x_B)\nonumber\\
&&+4\frac{2x_B-x_2}{x_2(x_2-x_1)}ln\bigl(\frac{x_2-x_B}{x_B}\bigr)
\theta (x_2-x_B) \nonumber\\
&&-4\frac{2x_B-x_1}{x_1(x_2-x_1)}ln\bigl(\frac{x_1-x_B}{x_B}\bigr)
\theta (x_1-x_B) \nonumber\\
&&+ln\left(\frac{m_q^2}{Q^2}\right)\biggl(-4\frac{2x_B-x_2}{x_2(x_2-x_1)}
\theta (x_2-x_B)+4\frac{2x_B-x_1}{x_1(x_2-x_1)}\theta (x_1-x_B)
\biggr) \nonumber\\
&&+(x_i\rightarrow -x_i)\biggr]+(x_B\rightarrow -x_B) .
\end{eqnarray}
The terms in front of the IR logarithms is nothing but the leading
logarithmic twist-three splitting functions.

We have finished with the calculation and now are in position
to demonstrate some interesting consequences of these results.
First of all, we find that the Burkhardt-Cottingham sum rule
is satisfied supporting the well known result that in the light cone
expansion there is no operator corresponding to the finite value
of the first moment. However, this assertion corresponds to the contribution
of local operators. In our approach there are operators
that can enter the first moments of $g_2$ as well as $g_1$, they are of
nonlocal origin: $\partial_{+}^{-1}G_{+\mu}\partial_{+}^{-1}
G_{+\nu}\partial_{+}^{-1}\widetilde G_{+\mu}$
and $\partial_{+}^{-1}G_{+\mu}\partial_{+}^{-1}\widetilde G_{+\nu}
\partial_{+}^{-1}G_{+\mu}$.
But as follows from the eqs.(13)-(15) their Wilson coefficients vanish:
\begin{equation}
\int \limits_{0}^{1}dx_Bg_2^{3gluon}(x_B)=0 .
\end{equation}
Moreover
\begin{equation}
\int \limits_{0}^{1}dx_Bg_1^{3gluon}(x_B)=0 .
\end{equation}
This equation seems to be quite obvious because the nonzero
value of the first moment of $g_1$ would lead to the violation
of the Burkhardt-Cottingham sum rule. This observation follows from
the facts arisen in the process of calculations. When we take the
convolution of gluon legs indices with the tensor structure of
$\widetilde N$-function in eq.(5) we extract the coefficients
corresponding to the polarized structure function $g_1$ ($\widetilde E_1$)
and to the sum $g_1+g_2$ ($\widetilde E_{1+2}$), not $g_1$ and $g_2$
separately. Therefore, the Burkhardt-Cottingham sum rule could not be
violated for $M_{1}(g_1)\not = 0$ if the first moment of the difference
$\widetilde E_{1+2}-\widetilde E_{1}$ equals zero. But we have the
situation when the first moment of each term vanishes ( eq.(17) ).

A new feature appears in the third moments which was absent in
the previous analysis of Bukhvostov, Kuraev and Lipatov:
a set of nonlocal operators:
\begin{eqnarray}
&&\!\!\!\!\!\!\!\!\!\!\!\!\int \limits_{0}^{1}dx_Bx_B^2g_2^
{3gluon}(x_B)\nonumber\\
&&\!\!\!\!\!\!\!\!\!\!\!\!=\langle Q_q^2\rangle N_f\frac{{\alpha}_s}
{4\pi}\Biggl[\left(\frac{5}{3}+\frac{2}{3}ln\left(\frac{m_q^2}{Q^2}
\right)\right)3\int dx_1dx_2\left(\frac{1}{x_1}+\frac{1}{x_2}\right)
N(x_1,x_2) \nonumber\\
&&\,\;\;\;\;\;\;\;\;\;\;\;+\left(-\frac{10}{3}-\frac{4}{3}ln\left(
\frac{m_q^2}{Q^2}\right)\right)6\int dx_1dx_2\frac{
\widetilde N(x_1,x_2)}{x_2-x_1} \nonumber\\
&&\;\;\;\;+\left(-\frac{15}{3}-\frac{6}{3}ln\left(\frac{m_q^2}{Q^2}
\right)\right)3\int dx_1dx_2\left(\frac{1}{x_1}-\frac{1}{x_2}\right)
\widetilde N(x_1,x_2)\Biggr]
\end{eqnarray}
and
\begin{eqnarray}
&&\!\!\!\!\!\!\!\!\!\!\!\!\!\!\!\!\!\!\!\!\!\int \limits_{0}^{1}dx_B
x_B^2g_1^{3gluon}(x_B)\nonumber\\
&&\!\!\!\!\!\!\!\!\!\!\!\!\!\!\!\!\!\!\!\!\!=\langle Q_q^2\rangle N_f
\frac{{\alpha}_s}{4\pi}\Biggl[\left(\frac{15}{3}+\frac{6}{3}ln\left(
\frac{m_q^2}{Q^2}\right)\right)6\int dx_1dx_2\frac{\widetilde N(x_1,x_2)}
{x_2-x_1} \nonumber\\
&&\;\;\;\;\;\;+\left(\frac{10}{3}+\frac{4}{3}ln\left(\frac{m_q^2}{Q^2}
\right)\right)3\int dx_1dx_2\left(\frac{1}{x_1}-\frac{1}{x_2}\right)
\widetilde N(x_1,x_2)\Biggr] .
\end{eqnarray}
Where the integrals are expressed through the bilocal gluon operators:
\begin{eqnarray}
&&3\int dx_1dx_2\left(\frac{1}{x_1}+\frac{1}{x_2}\right)2s_{\perp}^2
N(x_1,x_2) \nonumber\\
&&=-g(in^{-})^3s_{\perp}^{\nu}\int d{\mu}\epsilon (\mu )if^{abc}
\langle p,s|2G_{+\mu }^a(0)\widetilde G_{+\nu }^b(\mu n)G_{+\mu }^c
(\mu n)\nonumber\\
&&+\widetilde G_{+\mu }^a(0)G_{+\mu }^b(\mu n)G_{+\nu }^c(\mu n)
+G_{+\nu }^a(0)G_{+\mu }^b(\mu n)\widetilde G_{+\mu }^c(\mu n)|p,s\rangle
\end{eqnarray}
and
\begin{eqnarray}
&&6\int dx_1dx_2\frac{2s_{\perp}^2\widetilde N(x_1,x_2)}
{x_2-x_1} \nonumber\\
&&=-g(in^{-})^3s_{\perp}^{\nu}\int d{\mu}\epsilon (\mu )
if^{abc}\langle p,s|-2G_{+\mu }^a(0)\widetilde G_{+\nu }^b(\mu n)
G_{+\mu }^c(\mu n)\nonumber\\
&&-4\widetilde G_{+\mu }^a(0)G_{+\mu }^b(\mu n)G_{+\nu }^c(\mu n)
-G_{+\nu }^a(0)G_{+\mu }^b(\mu n)\widetilde G_{+\mu }^c(\mu n)|p,s\rangle
\end{eqnarray}
and
\begin{eqnarray}
&&3\int dx_1dx_2\left(\frac{1}{x_1}-\frac{1}{x_2}\right)2s_{\perp}^2
\widetilde N(x_1,x_2) \nonumber\\
&&=-g(in^{-})^3s_{\perp}^{\nu}\int d{\mu}\epsilon (\mu )if^{abc}
\langle p,s|G_{+\mu }^a(0)\widetilde G_{+\nu }^b(\mu n)G_{+\mu }^c
(\mu n)\nonumber\\
&&-\widetilde G_{+\mu }^a(0)G_{+\mu }^b(\mu n)G_{+\nu }^c(\mu n)
+2G_{+\nu }^a(0)G_{+\mu }^b(\mu n)\widetilde G_{+\mu }^c(\mu n)|p,s\rangle .
\end{eqnarray}
Of cause we can easily restore the gauge invariant result
substituting the path dependent exponents in the adjoint
representation of the colour group into the bilocal operators.

For transversely polarized nucleon the combination $g_1+g_2$
is measured experimentally:
\begin{eqnarray}
&&\int \limits_{0}^{1}dx_Bx_B^2(g_1^{3gluon}(x_B)
+g_2^{3gluon}(x_B))\nonumber\\
&&=\langle Q_q^2\rangle N_f\frac{{\alpha}_s}{4\pi}
\left(\frac{5}{3}+\frac{2}{3}ln\left(\frac{m_q^2}{Q^2}
\right)\right) \Biggl[3\int dx_1dx_2\left(\frac{1}{x_1}
+\frac{1}{x_2}\right)N(x_1,x_2) \nonumber\\
&&+6\int dx_1dx_2\frac{\widetilde N(x_1,x_2)}{x_2-x_1}
-3\int dx_1dx_2\left(\frac{1}{x_1}-\frac{1}{x_2}\right)
\widetilde N(x_1,x_2)\Biggr] .
\end{eqnarray}
So, we differ in a number of points with Ref. [8,{\bf a}]. First, the
evolution equation is written for the function $L(x_1,x_2)$ which
is some combination of functions $N(x_1,x_2)$ and $\widetilde N(x_1,x_2)$:
\begin{eqnarray}
&&L(x_1,x_2)=[2N(x_1,x_1-x_2)+2N(x_2,x_2-x_1)-\nonumber\\
&&\widetilde N(x_1,x_1-x_2)-\widetilde N(x_2,x_2-x_1)] .
\end{eqnarray}
This equality holds up to the overall numerical constant which is not
important for further discussion. So, we cannot express $N$ and/or
$\widetilde N$
separately through the function $L$, and therefore cannot compare
the twist-3 splitting functions. Constructing a combination of $L$'s
with different arguments we come to the equation:
\begin{eqnarray}
&&L(x_1,x_2)-L(x_1,x_1-x_2)-L(x_2,x_2-x_1)\nonumber\\
&&=4[N(x_1,x_2)-N(x_1,x_1-x_2)-N(x_2,x_2-x_1)] ,
\end{eqnarray}
which could not resolve the above problems. This equality takes
place because
of similar symmetry properties of $N$ and $L$ with respect to permutation
and change the sign of their arguments. But the most important
supplement is the appearance of the {\it nonlocal}\hspace{3mm}operators
which are present not only in the third moment
but also in the all highest moments of polarized structure functions.
This become a common feature for triple gluon contribution
as distinguished from the two gluon effect in $g_1$ widely
discussed earlier where a nonlocal operator appears
only in the first moment. But while in the last case it does not
affect the evolution equation, the twist-3 evolution equations must be
modified.

To summarise, we have calculated the coefficient functions of three-gluon
correlators, and have discussed their physical consequences.
The main results are:

(1) the Burkhardt-Cottingham sum rule is satisfied;

(2) the set of nonlocal operators appear in the moments of
structure functions.

The last problem deserves further investigation and will be considered
in the future publication.
\\
\\
{\Large\bf Acknowledgment}\\
\\
We thank E.A.Kuraev for useful conversation. This work was supported
by the International Science Foundation under grant RFEOOO and by the
Russian Foundation for Fundamental Investigation under the grant
\#93-02-3811.

\newpage
\pagestyle{empty}
\begin{figure}
\unitlength=2.80pt
\special{em:linewidth 0.4pt}
\linethickness{0.4pt}
\begin{picture}(154.00,131.00)
\put(43.56,39.50){\oval(3.01,3.01)[l]}
\put(43.56,37.49){\oval(3.01,3.01)[l]}
\put(44.06,38.49){\oval(2.01,1.00)[r]}
\put(44.06,36.49){\oval(2.01,1.00)[r]}
\put(43.56,35.48){\oval(3.01,3.01)[l]}
\put(43.56,33.47){\oval(3.01,3.01)[l]}
\put(44.06,34.48){\oval(2.01,1.00)[r]}
\put(43.56,31.46){\oval(3.01,3.01)[l]}
\put(43.56,29.46){\oval(3.01,3.01)[l]}
\put(44.06,30.46){\oval(2.01,1.00)[r]}
\put(44.06,28.45){\oval(2.01,1.00)[r]}
\put(43.56,27.45){\oval(3.01,3.01)[l]}
\put(43.56,25.44){\oval(3.01,3.01)[l]}
\put(44.06,26.44){\oval(2.01,1.00)[r]}
\put(44.06,32.47){\oval(2.01,1.00)[r]}
\put(63.56,39.50){\oval(3.01,3.01)[l]}
\put(63.56,37.49){\oval(3.01,3.01)[l]}
\put(64.06,38.49){\oval(2.01,1.00)[r]}
\put(64.06,36.49){\oval(2.01,1.00)[r]}
\put(63.56,35.48){\oval(3.01,3.01)[l]}
\put(63.56,33.47){\oval(3.01,3.01)[l]}
\put(64.06,34.48){\oval(2.01,1.00)[r]}
\put(63.56,31.46){\oval(3.01,3.01)[l]}
\put(63.56,29.46){\oval(3.01,3.01)[l]}
\put(64.06,30.46){\oval(2.01,1.00)[r]}
\put(64.06,28.45){\oval(2.01,1.00)[r]}
\put(63.56,27.45){\oval(3.01,3.01)[l]}
\put(63.56,25.44){\oval(3.01,3.01)[l]}
\put(64.06,26.44){\oval(2.01,1.00)[r]}
\put(64.06,32.47){\oval(2.01,1.00)[r]}
\put(53.56,39.50){\oval(3.01,3.01)[l]}
\put(53.56,37.49){\oval(3.01,3.01)[l]}
\put(54.06,38.49){\oval(2.01,1.00)[r]}
\put(54.06,36.49){\oval(2.01,1.00)[r]}
\put(53.56,35.48){\oval(3.01,3.01)[l]}
\put(53.56,33.47){\oval(3.01,3.01)[l]}
\put(54.06,34.48){\oval(2.01,1.00)[r]}
\put(53.56,31.46){\oval(3.01,3.01)[l]}
\put(53.56,29.46){\oval(3.01,3.01)[l]}
\put(54.06,30.46){\oval(2.01,1.00)[r]}
\put(54.06,28.45){\oval(2.01,1.00)[r]}
\put(53.56,27.45){\oval(3.01,3.01)[l]}
\put(53.56,25.44){\oval(3.01,3.01)[l]}
\put(54.06,26.44){\oval(2.01,1.00)[r]}
\put(54.06,32.47){\oval(2.01,1.00)[r]}
\put(65.50,61.00){\oval(3.00,2.00)[lt]}
\put(65.50,63.00){\oval(3.00,2.00)[rb]}
\put(68.50,63.00){\oval(3.00,2.00)[lt]}
\put(68.50,65.00){\oval(3.00,2.00)[rb]}
\put(71.50,65.00){\oval(3.00,2.00)[lt]}
\put(71.50,67.00){\oval(3.00,2.00)[rb]}
\put(74.50,67.00){\oval(3.00,2.00)[lt]}
\put(74.50,69.00){\oval(3.00,2.00)[rb]}
\put(77.50,69.00){\oval(3.00,2.00)[lt]}
\put(77.50,71.00){\oval(3.00,2.00)[rb]}
\put(30.50,69.00){\oval(3.00,2.00)[rt]}
\put(33.50,69.00){\oval(3.00,2.00)[lb]}
\put(30.50,71.00){\oval(3.00,2.00)[lb]}
\put(33.50,67.00){\oval(3.00,2.00)[rt]}
\put(36.50,67.00){\oval(3.00,2.00)[lb]}
\put(36.50,65.00){\oval(3.00,2.00)[rt]}
\put(39.50,65.00){\oval(3.00,2.00)[lb]}
\put(39.50,63.00){\oval(3.00,2.00)[rt]}
\put(42.50,63.00){\oval(3.00,2.00)[lb]}
\put(42.50,61.00){\oval(3.00,2.00)[rt]}
\put(53.56,55.50){\oval(3.01,3.01)[l]}
\put(53.56,53.49){\oval(3.01,3.01)[l]}
\put(54.06,54.49){\oval(2.01,1.00)[r]}
\put(54.06,52.49){\oval(2.01,1.00)[r]}
\put(53.56,51.48){\oval(3.01,3.01)[l]}
\put(53.56,49.47){\oval(3.01,3.01)[l]}
\put(54.06,50.48){\oval(2.01,1.00)[r]}
\put(53.56,47.46){\oval(3.01,3.01)[l]}
\put(53.56,45.46){\oval(3.01,3.01)[l]}
\put(54.06,46.46){\oval(2.01,1.00)[r]}
\put(54.06,44.45){\oval(2.01,1.00)[r]}
\put(53.56,43.45){\oval(3.01,3.01)[l]}
\put(53.56,41.44){\oval(3.01,3.01)[l]}
\put(54.06,42.44){\oval(2.01,1.00)[r]}
\put(54.06,48.47){\oval(2.01,1.00)[r]}
\put(54.06,40.45){\oval(2.01,1.00)[r]}
\put(53.56,59.49){\oval(3.01,3.01)[l]}
\put(54.06,58.49){\oval(2.01,1.00)[r]}
\put(53.56,57.48){\oval(3.01,3.01)[l]}
\put(54.06,56.48){\oval(2.01,1.00)[r]}
\put(69.00,70.00){\makebox(0,0)[cc]{$\beta$}}
\put(39.00,70.00){\makebox(0,0)[cc]{$\alpha$}}
\put(64.00,21.00){\makebox(0,0)[cc]{$\sigma$}}
\put(54.00,21.00){\makebox(0,0)[cc]{$\nu$}}
\put(44.00,21.00){\makebox(0,0)[cc]{$\mu$}}
\put(154.00,106.00){\makebox(0,0)[cc]{\bf (a)}}
\put(154.00,47.00){\makebox(0,0)[cc]{\bf (b)}}
\put(100.00,47.00){\makebox(0,0)[cc]{$ \left( \begin{array}{c}  permutation  \\
 of \mbox{ } gluon \mbox{ } legs,  \\  \alpha \rightarrow \beta  \end{array}
\right) $}}
\put(100.00,106.00){\makebox(0,0)[cc]{$ \left( \begin{array}{c}  permutation
\\  of \mbox{ } gluon \mbox{ } legs,  \\  \alpha \rightarrow \beta  \end{array}
\right) $}}
\put(73.33,47.00){\makebox(0,0)[cc]{+}}
\put(136.00,79.00){\makebox(0,0)[cc]{$\left[ \begin{array}{c} opposite \\
direction \hspace{2mm} of \\quark \hspace{2mm }line \end{array} \right]$}}
\put(110.67,79.00){\makebox(0,0)[cc]{+}}
\put(64.00,61.00){\vector(0,-1){11.00}}
\put(64.00,50.00){\line(0,-1){9.00}}
\put(44.00,41.00){\vector(0,1){11.00}}
\put(44.00,52.00){\line(0,1){9.00}}
\put(44.00,61.00){\vector(1,0){6.00}}
\put(54.00,61.00){\vector(1,0){6.00}}
\put(54.00,41.00){\vector(-1,0){6.00}}
\put(64.00,41.00){\vector(-1,0){6.00}}
\put(50.00,61.00){\line(1,0){4.00}}
\put(60.00,61.00){\line(1,0){4.00}}
\put(58.00,41.00){\line(-1,0){4.00}}
\put(48.00,41.00){\line(-1,0){4.00}}
\put(43.56,98.50){\oval(3.01,3.01)[l]}
\put(43.56,96.49){\oval(3.01,3.01)[l]}
\put(44.06,97.49){\oval(2.01,1.00)[r]}
\put(44.06,95.49){\oval(2.01,1.00)[r]}
\put(43.56,94.48){\oval(3.01,3.01)[l]}
\put(43.56,92.47){\oval(3.01,3.01)[l]}
\put(44.06,93.48){\oval(2.01,1.00)[r]}
\put(43.56,90.46){\oval(3.01,3.01)[l]}
\put(43.56,88.46){\oval(3.01,3.01)[l]}
\put(44.06,89.46){\oval(2.01,1.00)[r]}
\put(44.06,87.45){\oval(2.01,1.00)[r]}
\put(43.56,86.45){\oval(3.01,3.01)[l]}
\put(43.56,84.44){\oval(3.01,3.01)[l]}
\put(44.06,85.44){\oval(2.01,1.00)[r]}
\put(44.06,91.47){\oval(2.01,1.00)[r]}
\put(63.56,98.50){\oval(3.01,3.01)[l]}
\put(63.56,96.49){\oval(3.01,3.01)[l]}
\put(64.06,97.49){\oval(2.01,1.00)[r]}
\put(64.06,95.49){\oval(2.01,1.00)[r]}
\put(63.56,94.48){\oval(3.01,3.01)[l]}
\put(63.56,92.47){\oval(3.01,3.01)[l]}
\put(64.06,93.48){\oval(2.01,1.00)[r]}
\put(63.56,90.46){\oval(3.01,3.01)[l]}
\put(63.56,88.46){\oval(3.01,3.01)[l]}
\put(64.06,89.46){\oval(2.01,1.00)[r]}
\put(64.06,87.45){\oval(2.01,1.00)[r]}
\put(63.56,86.45){\oval(3.01,3.01)[l]}
\put(63.56,84.44){\oval(3.01,3.01)[l]}
\put(64.06,85.44){\oval(2.01,1.00)[r]}
\put(64.06,91.47){\oval(2.01,1.00)[r]}
\put(53.56,98.50){\oval(3.01,3.01)[l]}
\put(53.56,96.49){\oval(3.01,3.01)[l]}
\put(54.06,97.49){\oval(2.01,1.00)[r]}
\put(54.06,95.49){\oval(2.01,1.00)[r]}
\put(53.56,94.48){\oval(3.01,3.01)[l]}
\put(53.56,92.47){\oval(3.01,3.01)[l]}
\put(54.06,93.48){\oval(2.01,1.00)[r]}
\put(53.56,90.46){\oval(3.01,3.01)[l]}
\put(53.56,88.46){\oval(3.01,3.01)[l]}
\put(54.06,89.46){\oval(2.01,1.00)[r]}
\put(54.06,87.45){\oval(2.01,1.00)[r]}
\put(53.56,86.45){\oval(3.01,3.01)[l]}
\put(53.56,84.44){\oval(3.01,3.01)[l]}
\put(54.06,85.44){\oval(2.01,1.00)[r]}
\put(54.06,91.47){\oval(2.01,1.00)[r]}
\put(65.50,120.00){\oval(3.00,2.00)[lt]}
\put(65.50,122.00){\oval(3.00,2.00)[rb]}
\put(68.50,122.00){\oval(3.00,2.00)[lt]}
\put(68.50,124.00){\oval(3.00,2.00)[rb]}
\put(71.50,124.00){\oval(3.00,2.00)[lt]}
\put(71.50,126.00){\oval(3.00,2.00)[rb]}
\put(74.50,126.00){\oval(3.00,2.00)[lt]}
\put(74.50,128.00){\oval(3.00,2.00)[rb]}
\put(77.50,128.00){\oval(3.00,2.00)[lt]}
\put(77.50,130.00){\oval(3.00,2.00)[rb]}
\put(30.50,128.00){\oval(3.00,2.00)[rt]}
\put(33.50,128.00){\oval(3.00,2.00)[lb]}
\put(30.50,130.00){\oval(3.00,2.00)[lb]}
\put(33.50,126.00){\oval(3.00,2.00)[rt]}
\put(36.50,126.00){\oval(3.00,2.00)[lb]}
\put(36.50,124.00){\oval(3.00,2.00)[rt]}
\put(39.50,124.00){\oval(3.00,2.00)[lb]}
\put(39.50,122.00){\oval(3.00,2.00)[rt]}
\put(42.50,122.00){\oval(3.00,2.00)[lb]}
\put(42.50,120.00){\oval(3.00,2.00)[rt]}
\put(69.00,129.00){\makebox(0,0)[cc]{$\beta$}}
\put(39.00,129.00){\makebox(0,0)[cc]{$\alpha$}}
\put(64.00,80.00){\makebox(0,0)[cc]{$\sigma$}}
\put(54.00,80.00){\makebox(0,0)[cc]{$\nu$}}
\put(44.00,80.00){\makebox(0,0)[cc]{$\mu$}}
\put(73.33,106.00){\makebox(0,0)[cc]{+}}
\put(64.00,120.00){\vector(0,-1){11.00}}
\put(64.00,109.00){\line(0,-1){9.00}}
\put(44.00,100.00){\vector(0,1){11.00}}
\put(44.00,111.00){\line(0,1){9.00}}
\put(54.00,100.00){\vector(-1,0){6.00}}
\put(64.00,100.00){\vector(-1,0){6.00}}
\put(58.00,100.00){\line(-1,0){4.00}}
\put(48.00,100.00){\line(-1,0){4.00}}
\put(44.00,120.00){\vector(1,0){11.00}}
\put(55.00,120.00){\line(1,0){9.00}}
\end{picture}
\caption{The lowest order diagrams for the coefficient function}
\end{figure}
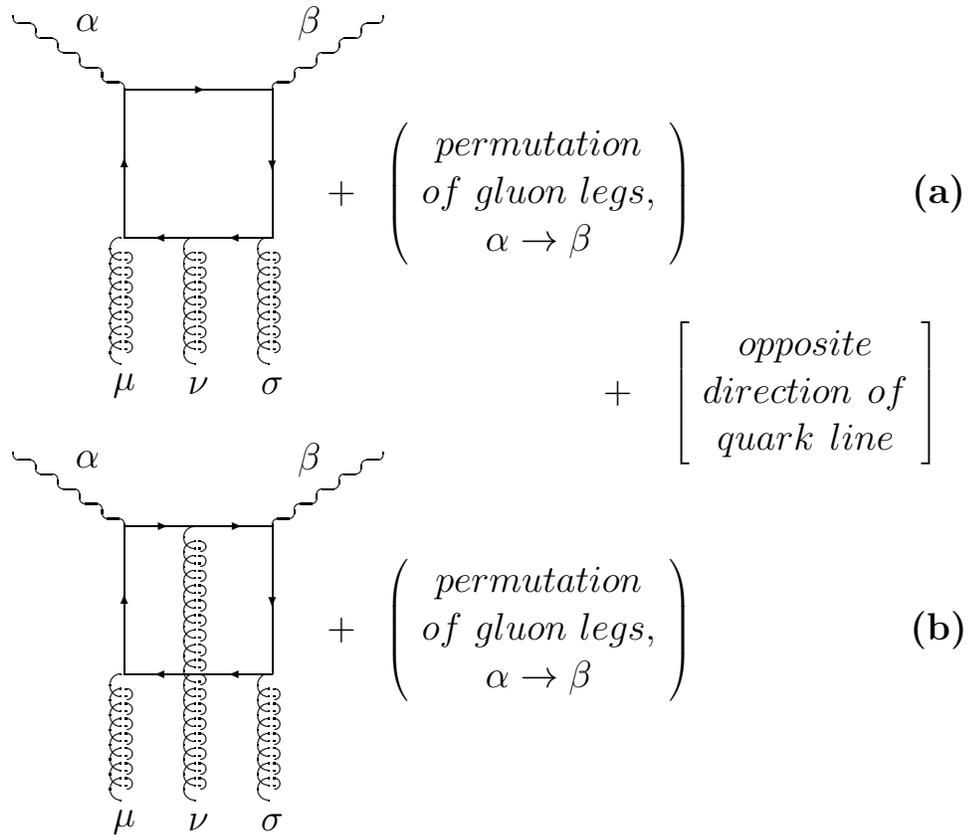
\end{document}